\documentstyle[aasms4]{article}

\begin{document}

\title{Scalar dark matter in spiral galaxies.}
\author{F. Siddhartha Guzm\'an\footnote{E-mail: siddh@fis.cinvestav.mx},
        Tonatiuh Matos }
\affil{{\it Departamento de F\'{\i}sica, Centro de Investigaci\'on y de
       Estudios Avanzados del IPN, A.P. 14-740, 07000 M\'exico D.F.,
       MEXICO.}}
\author{Hugo Villegas-Brena}
\affil{{\it Instituto de Ciencias Nucleares,
       Universidad Nacional Aut\'{o}noma de M\'{e}xico,
       A.P. 70-543, 04510 M\'{e}xico D.F., MEXICO.}}
\authoremail{siddh@fis.cinvestav.mx}

\begin{abstract}
An exact, axially symmetric solution to the Einstein-Klein-Gordon field
equations is employed to model the dark matter in spiral galaxies. The
extended rotation curves from a previous analysis are used to fit the
model and a very good agreement is found. It is argued that, although our
model possesses three parameters to be fitted, it is better than the
non-relativistic alternatives in the sense that it is not of a
phenomenological nature, since the dark matter would consist entirely of a
scalar field.
\end{abstract}

{\bf Key words:} galaxies: haloes -- cosmology: dark matter.

\section{Introduction.}

Since the pioneering works by Oort and Zwicky, back in the 1930's
(\cite{Oort}; \cite{Zwicky}), the existence of dark matter in the Universe
has been firmly established by astronomical observations at very different
length-scales, ranging from the neighbourhood of the Solar System to the
clusters of galaxies. What it means is that a large fraction of the mass
needed to produce, within the framework of Newtonian mechanics, the observed
dynamical effects in all these very different systems, is not seen. This
puzzle has stimulated the exploration of lots of proposals, and very
imaginative explanations have been put forward, from exotic matter to
nonrelativistic modifications of Newtonian dynamics.\\

In particular, the measurement of rotation curves (RC) in galaxies shows
that the coplanar orbital motion of gas in the outer parts of these galaxies
keeps a more or less constant velocity up to several luminous radii. The
discrepancy arises when one applies the usual Newtonian dynamics to the
observed luminous matter and gas, since then the circular velocity should
decrease as we move outwards. The most widely accepted explanation is that
of a spherical halo of dark matter, its nature being unknown, which would
surround the galaxy and account for the missing mass needed to produce the
flat RC. Another possibility, considered much less often, is the so called
Modified Newtonian Dynamics (MOND), which was put forward by Milgrom
(1983);
in this model, the usual second Newton law would broke at ``small''
accelerations, as compared to some (in principle universal) acceleration
parameter, $a_0$. Although it seems to provide for a very good
phenomenological description of the RC, it lacks, at least until now, a more
sound theoretical basis.\\

Our aim here is to give another explanation to the dark matter problem in
spiral galaxies, this time using a fully relativistic approach, and making
use of the well known scalar fields. Combining an exact solution of the
Einstein-Klein-Gordon field equations with the observations from luminous
matter and gas, we are able to reproduce the flat extended RC of spiral
galaxies. Scalar fields are the simplest generalization of General
Relativity and they can be introduced on very fundamental grounds, as in
the case of the Brans-Dicke, the Kaluza-Klein and the Super-Strings
theories. They appear also in cosmological models, like inflation, and in
general in all modern unifying theories. Recently, it has been suggested
(\cite{Cho}; \cite{peebles}) that a massive scalar field could account for
the dark matter at cosmological scales, and a previous work
(\cite{MatGuzb}) has shown a preliminary analysis in the
context of spiral galaxies. This is encouraging since it makes models like
the one proposed here to seem more plausible.\\

The paper is organized as follows: in \S \ref{sec:metric} we introduce the
field equations and the explicit solution for an axially symmetric
configuration; in \S \ref{sec:model} the geodesic equations are written and
the model for the dark matter in a spiral galaxy is introduced; section \S \ref
{sec:results} gives the main results concerning the fitting of the model to
the observations, and in \S \ref{sec:end} some concluding remarks are done.
Finally, two appendices describe some geometrical properties of our metric
and make a brief conceptual comparison with the dark halo and the MOND
hypotheses.\\

\section{The field equations and their solution.}

\label{sec:metric}

As mentioned in the Introduction, scalar fields appear in a natural way
within the framework of unifying theories. As example we mention
Kaluza-Klein (KK) and Super-Strings (SS), where the scalar field 
appears in the effective action after dimensional reduction. Let us begin
with the most general scalar-tensor theory of gravity, as given by the
action:

\begin{equation}
\tilde{{\cal S}}=\int d^{4}x\sqrt{-\tilde{g}}\left( -\frac{F}{\kappa _{0}}%
\tilde{R}+(\tilde{\nabla}\Phi )^{2}G+\tilde{V}(\Phi )\right) .
\end{equation}

\noindent where $F,$ \ $\tilde{V}$ and $G$ are functions of $\Phi $ only, $%
\tilde{g}$ is the determinant of the metric $\tilde{g}_{\mu\nu}$ and $\kappa
_{0}=16\pi G/c^{4}$.
We can perform a conformal transformation to some other frame by means of the
redefinitions (\cite{damour2}; \cite{Frolov}),

\begin{equation}
g_{\mu \nu }=F\tilde{g}_{\mu \nu },
\end{equation}

\begin{equation}
\phi =\frac{1}{\sqrt{2}}\int d\Phi \left[ \frac{3}{2}\left( \frac{F^{\prime }%
}{F}\right) ^{2}+\frac{G}{F}\right] ^{1/2},
\end{equation}
where the prime denotes $d/d\Phi $. The action then takes the form:

\begin{equation}
S=\int d^{4}x\sqrt{-g}[-\frac{R}{\kappa _{0}}+2(\nabla \phi )^{2}+V(\phi )],
\label{eq:action}
\end{equation}

\noindent where $R$ is the four dimensional scalar curvature and $g$ the
determinant of the metric $g_{\mu \nu }$. $\phi $ and $V(\phi )$ repectively are
the scalar field and the scalar potential in the new frame, and all
the other quantities are also calculated using the new metric. The choice of the
Einstein frame, implicit in this form of the action, is made because the field
equations are more easily solved in this frame. As is well known, since the
coupling of a scalar field with gravity is defined up to a conformal
transformation, there is some ambiguity in regarding a specific frame as the
``physical'' frame (see, e.g., \cite {Damour}). We shall consider in
detail what happens both to the action above and to the equations of
motion in section \ref{sec:model}.\\

The next step is to decide what kind of potential $V(\phi )$ is the most
convenient for modeling a galaxy. Let us reason as follows. It is known that
the energy density of the dark matter in the halo of the galaxies goes like $%
1/r^{2}$. The energy momentum tensor of the scalar field is basically the
sum of quadratic terms of the scalar field derivatives plus the scalar
potential, $i.e.$, $T_{\mu \nu }\sim \phi _{,\mu }\phi _{,\nu }+V(\phi )\sim
1/r^{2}.$ If we assume that the term $\phi _{,\mu }\phi _{,\nu }\sim 1/r^{2}$
as well as the term $V(\phi )\sim 1/r^{2}$, from the first assumption we
infer that $\phi \sim \ln (r)$, and from the second assumption we arrive at $%
V(\phi )\sim \exp (-2\phi )$. So, in what follows we will take the potential 
$V(\phi )=\Lambda \exp (-2\alpha \phi )$ and we shall consider the four
dimensional action:

\[
S=\int d^{4}x\sqrt{-g}[-\frac{R}{\kappa _{0}}+2(\nabla \phi
)^{2}+e^{-2\alpha \phi }\Lambda ]. 
\]

For the time being, from this very general setting we shall look for an
exact solution to the field equations that will serve us as a model for a
spiral galaxy. Now, since the velocity of the gas and the red shift
measurements in a galaxy are made over the equatorial plane, it is
reasonable to impose axial symmetry on the solution we are looking for, in
contrast with the usual spherical dark halo profile used in most studies
(\cite{Begeman}). Moreover, the fact that a substantial
amount of the total mass in these galaxies is in the form of dark matter
suggests that, in a first approximation, the observed baryonic mass (both
stars and gas) will not contribute significantly to the total energy density
of the system, at least in the region outside the luminous disk; instead,
the scalar matter will determine the space-time curvature, and the material
particles will move on geodesics determined (almost) by the energy density
of the scalar field. Finally, the RC exhibits a constant velocity of the
order of 100-350 km/s, which, compared to the velocity of light, clearly
allows one to consider the galaxy as a static system.\\

With the above simplifying assumptions, the most general axially symmetric,
static metric can be written as:

\begin{equation}  \label{eq:metric}
ds^2 = \frac{1}{f}[e^{2k}(dz d\overline{z})+W^2d\varphi^2]-fc^2dt^2,
\end{equation}

\noindent where $z:=\rho+i\zeta$, the bar means complex conjugation, and the
real valued functions $f,W$ and $k$ depend only on $z$ and $\overline{z}$
(or equivalently on $\rho$ and $\zeta$).\\

After varying the action in equation~(\ref{eq:action}), one obtains the
following field equations:

\[
\phi _{;\mu }^{;\mu }-\frac{1}{4}\frac{dV}{d\phi }=0, 
\]

\begin{equation}
R_{\mu \nu }=\kappa _{0}[2\phi _{,\mu }\phi _{{,\nu }}+\frac{1}{2}g_{\mu \nu
}V(\phi )],  \label{eq:feqs}
\end{equation}

\noindent which are the Klein-Gordon and Einstein field equations,
respectively, and $\mu,\nu=0,1,2,3$. A very powerful technique, known as the
harmonic maps ansatz, can be employed to find families of solutions to the
equations~(\ref{eq:feqs}), starting with the metric~(\ref{eq:metric}). The
details can be found in Matos (1989; 1994; 1995) and Guzm\'an \&
Matos (1999), so we shall only describe it very briefly here.\\

\subsection{The harmonic maps ansatz.}

In a few words, the main idea behind the method is to re-parameterize the
functions in metric~(\ref{eq:metric}) with convenient auxiliary functions
which will obey a generalization of the Laplace equation, along with some
consistency relationships; the latter are usually quite difficult to
fulfill, and great care and intuition must be taken in order to get a system
of equations both workable with and interesting enough. In this case we
shall take $f=e^{\lambda }$, and assume that $f$ and $\phi $ are functions
of $W$, which in turn is a function of $z$ and $\overline{z}$ alone. After
lengthy but straightforward calculations one is left with the system:

\begin{eqnarray}
\hat{\Delta}\lambda &=&\kappa _{0}\sqrt{-g}V(\phi ), 
\nonumber \\
2\hat{\Delta}\phi &=&-\frac{1}{4}\sqrt{-g}\frac{dV}{d\phi },  \nonumber \\
W_{,z\overline{z}} &=&\frac{1}{2}\kappa _{0}\sqrt{-g}V(\phi ), \nonumber\\
k_{,z} &=&\frac{W_{,zz}}{2W_{,z}}+\frac{W}{4}\lambda _{,z}^{2}W_{,z}+\kappa
_{0}W\phi _{,z}^{2}W_{,z},  \label{eq:set}
\end{eqnarray}

\noindent and a similar equation for $k_{,\overline{z}}$, replacing $z$ by $%
\overline{z}$. The symbol $\hat{\Delta}$ stands for a generalized Laplace
operator, such that for every function $h(z,\overline{z})$, $\hat{\Delta}h
:= (Wh_{,z})_{,\overline{z}}+(Wh_{,\overline{z}})_{,z}$.

\subsection{The model for the galaxy.}

Regarding the set of equations~(\ref{eq:set}), it can be noted that the last
equation and its complex conjugate are integrable once the functions $%
\lambda ,\phi $ and $W$ have been integrated. The first three equations,
however, are highly coupled since $\sqrt{-g}=W\exp {(2k-\lambda )}/2$;
moreover, the operator $\hat{\Delta}$ itself contains $W$. In spite of this,
we have been able to obtain a not too restrictive solution, which can be
written as:

\begin{eqnarray}
\lambda &=&\ln {M}+\ln {f_{0}},  \nonumber  \nonumber \\
\phi &=&\phi _{0}+\frac{1}{2\sqrt{\kappa _{0}}}\ln {M},  \nonumber \\
V &=&-\frac{4f_{0}}{\kappa _{0}M}, \nonumber\\
e^{2k} &=&M_{,z\overline{z}}M,  \label{eq:sol}
\end{eqnarray}

\noindent where $f_{0}$ and $\phi _{0}$ are integration constants with

\begin{equation}
e^{-2\sqrt{\kappa _{0}}\phi _{0}}=4f_{0}\Lambda /\kappa _{0}  \label{const}
\end{equation}
and $M\equiv W$ is, as stated before, a function of $z$ and $\overline{z}$,
restricted only by the condition

\begin{equation}  \label{eq:condM}
MM_{,z\overline{z}} = M_{,z}M_{,\overline{z}},
\end{equation}

\noindent but otherwise arbitrary. Observe that $\Lambda $ and $%
\kappa _{0}$ are fundamental constants of the theory but $f_{0}$ and $\phi_{0}$
 are integration constants, $i.e.$, they are different for each
space-time (each galaxy), fulfilling the relation (\ref{const}). The values
of these constants will determine the characteristics of a particular
galaxy.\\

We shall take the solution~(\ref{eq:sol}) as the general relativistic
description of the galaxy, considering a particular choice of $M$, namely, $%
M=z\overline{z}/r_0$, where $r_0$ is a constant with dimensions of
length.\\

\section{Geodesic motion along the equatorial plane.}

\label{sec:model}

Since the particles compossing the gas from which observations arise are
small compared to the whole galaxy, they can be
considered as test particles moving on the background metric~(\ref
{eq:metric}). Therefore, the next step is to study the geodesic motion of
test particles along the equatorial plane. From metric~(\ref{eq:metric}) we
can write, for material particles:

\begin{equation}
\frac{1}{f}\left[ e^{2k}\left( \left( \frac{d\rho }{d\tau }\right)
^{2}+\left( \frac{d\zeta }{d\tau }\right) ^{2}\right) +W^{2}\left( \frac{%
d\varphi }{d\tau }\right) ^{2}\right] -fc^{2}\left( \frac{dt}{d\tau }\right)
^{2}=c^{2}.  \label{eq:geod}
\end{equation}
As the solution is axially symmetric and static, there will be two constants
of motion, namely, the angular momentum per unit mass,

\[
B=\frac{W^{2}}{f}\frac{d\varphi }{d\tau }, 
\]

\noindent and the total energy of the test particle,

\[
A = fc^2\frac{dt}{d\tau}, 
\]

\noindent where $\tau$ is the proper time of the test particle. In order to
obtain useful information from these constants, it is convenient to write
the line element as:

\begin{eqnarray}
ds^{2} &=&\left[ \frac{1}{fc^{2}}\left( e^{2k}(\dot{\rho}^{2}+\dot{\zeta}%
^{2})+W^{2}\dot{\varphi}^{2}\right) -f\right] c^{2}dt^{2}  
\label{eq:interval} \\
&=&\left( \frac{v^{2}}{c^{2}}-f\right) c^{2}dt^{2},
\end{eqnarray}

\noindent since the squared three-velocity, $v^2$, is given by:

\[
v^{2}=g_{ab}v^{a}v^{b}=\frac{e^{2k}}{f}(\dot{\rho}^{2}+\dot{\zeta}^{2})+%
\frac{W^{2}}{f}\dot{\varphi}^{2}
\]

\noindent where $a,b=1,2,3$. On the other hand, for a material freely
falling observer (i.e., an observer in geodesic motion) we must have $ds^2 =
-c^2d\tau^2$, and equating this expression with equation~(\ref{eq:interval}%
), we can arrive to:

\begin{equation}
A^{2}=\frac{c^{4}f^{2}}{f-v^{2}/c^{2}}.  \label{eq:A}
\end{equation}
If we now identify the equatorial plane of the galaxy with the plane $\zeta
=0$, the geodesic equation~(\ref{eq:geod}) reduces to the following, after
inserting the constants of motion $A$ and $B$:

\begin{equation}
\frac{1}{f}e^{2k}\left(\frac{d\rho}{d\tau}\right)^2+\frac{B^2f}{W^2} -\frac{%
A^2}{c^2 f} = c^2.
\end{equation}

\noindent This equation describes the motion of test particles in the
equatorial plane of the galaxy, and in every particular trajectory the
constants $A$ and $B$ remain so, i.e., constant. However, a key point here
is that, shall we change of trajectory, the values for $A$ and $B$ will
change accordingly for the new trajectory. Since the RC give an average of
the circular velocity of particles in the galaxy, we shall consider circular
orbits only, for which $\dot{\rho} = 0$, so that $v$ in equation~(\ref
{eq:interval}) can be identified with $v_c\equiv v_{circular}$. From this,
an expression for $B$ in terms of $v^2$ can be written down:

\[
B^2 = \frac{v^2}{f-v^2/c^2}\frac{W^2}{f}, 
\]

\noindent and since, as stated before, $v \ll c$, this gives:

\begin{equation}  \label{eq:B}
B^2 \cong v^2\frac{W^2}{f^2}.
\end{equation}

\noindent Using now the form of $f$ given by equations~(\ref{eq:sol}), one
arrives at the following remarkably simple relation between $v$ and $B$:

\begin{equation}
v_{DM}=f_{0}B,  \label{eq:vdm}
\end{equation}

\noindent where we have written $v_{DM}$ instead of $v$ to stress the fact
that this velocity for test particles is due to the scalar dark matter.
Formula (\ref{eq:vdm}) is the main result of this model, it states the way
how the circular velocity due to the dark matter is determined by the
angular momentum from each orbit. It is remarkable that formula (\ref{eq:vdm}%
) is invariant under conformal transformation $g_{\mu \nu }=F\tilde{g}_{\mu
\nu }$ of the metric, $i.e.$, this formula is valid for the metrics $g_{\mu
\nu }$ and $\tilde{g}_{\mu \nu }.$\\

In order to gain some insight into the physical meaning of the solution~(\ref
{eq:sol}), we write it in Boyer-Lindquist-like coordinates (Schwarzschild
coordinates) $(r,\theta ,\varphi )$, related to $(\rho ,\zeta ,\varphi )$ by 
$\rho =\sqrt{r^{2}+b^{2}}\sin {\theta }$, $\zeta =r\cos {\theta }$; metric~(%
\ref{eq:metric}) then reads:

\begin{equation}
ds^{2}=\frac{1+b^{2}\cos ^{2}{\theta }/r^{2}}{f_{0}r_{0}}\left( \frac{dr^{2}%
}{1+b^{2}/r^{2}}+r^{2}d\theta ^{2}\right) +\frac{r^{2}+b^{2}\sin ^{2}{\theta 
}}{f_{0}r_{0}}d\varphi ^{2}-f_{0}c^{2}\frac{r^{2}+b^{2}\sin ^{2}{\theta }}{%
r_{0}}dt^{2}.  \label{eq:metricBL}
\end{equation}

\noindent On the other hand, the effective energy density $\mu_{DM}$ is
given by:

\begin{equation}
\mu _{DM}=\frac{1}{2}V(\phi )=-\frac{2f_{0}r_{0}}{\kappa
_{0}(r^{2}+b^{2}\sin ^{2}{\theta )}}.  \label{eq:mudm}
\end{equation}

The fact that this energy density is negative does not constitute a serious
drawback since, as mentioned before, we most perform a conformal
transformation of the metric (\ref{eq:metric}) in order to obtain the action
corresponding to a theory with a more physical interpretation.\\

To be able to obtain more quantitative results and to compare this model
with the most usual approaches, one further assumption must be made,
regarding the constant of motion $B$. The observed luminous matter in a
galaxy behaves in accord to Newtonian dynamics to a good approximation, so
that its angular momentum per unit mass will be $B=v_{L}\times D$, where $%
v_{L}$ is the contribution of the luminous matter, and $D$ is the interval
from the metric as written in equation~(\ref{eq:metricBL}); as we are in the
equatorial plane, on circular orbits and at one particular instant, we have $%
d\rho =d\zeta =dt=0$, and it is easy to check that $D=\sqrt{%
(r^{2}+b^{2})/f_{0}}$. It is now reasonable to substitute this value for $B$
in equation~(\ref{eq:vdm}), since the expression for $v_{DM}$ represents the
velocity of test particles due to the presence of the scalar field; we get:

\begin{equation}
v_{DM}=f_{0}v_{L}\sqrt{(r^{2}+b^{2})/f_{0}},  \label{eq:vdmb}
\end{equation}

\noindent Noting that the total kinetic energy will be well approximated by
the sum of the individual contributions, i.e., $
1/2 mv_C^2 \simeq 
1/2 mv_{L}^2+
1/2 mv_{gas}^2+ 
mv_{DM}^2$, we arrive at
the
final form of the velocity along circular trajectories in the equatorial
plane of the galaxy:

\begin{equation}
v_{C}^{2}=v_{L}^{2}(1+f_{0}(r^{2}+b^{2}))+v_{gas}^{2},  \label{eq:vctot}
\end{equation}

\noindent where the constants $f_0$ and $b$ will be parameters to adjust to
the observed RC. To this end we shall proceed as follows: we take the
photometric and RC data for 6 spiral galaxies from Begeman et al. (1991)
and Kent (1987), as listed in Table~\ref{tab:sample}.
This data is fitted using
a non-linear least squares routine adding a third parameter, namely, the
usual mass-luminosity ratio $M/L$, which is taken to be constant in each
particular galaxy; when there exist disk and bulge observations, two $M/L$
ratios are assumed. The total luminous mass at a distance $r$ from the
center of the galaxy will be $M_L(r) = (M/L) \times L(r)$, i.e.:

\begin{equation}  \label{eq:vlum}
v_{L}^2(r) = \frac{GM_L(r)}{r}.
\end{equation}

\noindent Combining equations~(\ref{eq:vctot}) and~(\ref{eq:vlum}), and
including the 21 cm data from gas, we are led to:

\begin{equation}
v_{C}^{2}(r)=\frac{GM_{L}(r)}{r}(1+f_{0}(r^{2}+b^{2}))+v_{gas}^{2}.
\label{eq:modeltofit1}
\end{equation}

In the next section~\ref{sec:results} we shall show how this compares with the
actual observational data. It can also be useful to compare this derivation with
more common explanations for the RC, two of which are briefly described in appendix
\ref{sec:dhmond}.

\section{Results.}

\label{sec:results}

\begin{figure*}
\label{fig:fits}
\centerline{ \epsfxsize=12cm \epsfbox{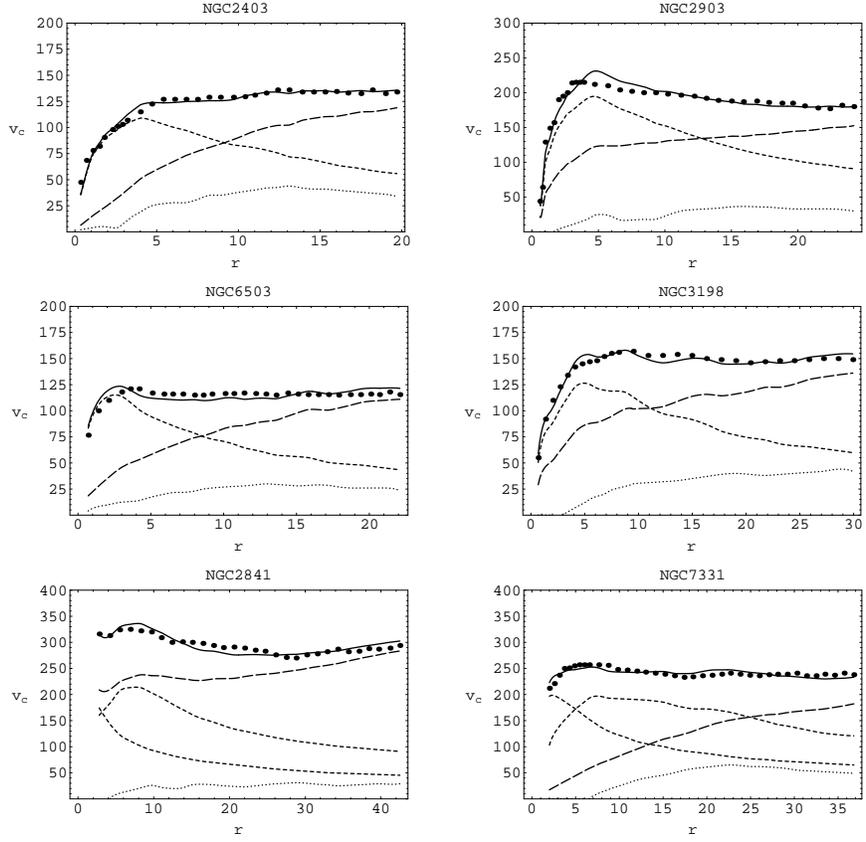}}
\caption{
Curves of $v_C$ (km/s) vs $r$ (kpc) for the observational data (dots) and
the fits from equation (\ref{eq:modeltofit1}) (solid). Also shown are the
individual contributions from luminous matter (short dashed), the dilaton
(long dashed) and gas (dotted). In the case of NGC 2841 and NGC 7331, the lower
short dashed curve represents the contribution from the bulge of these galaxies.
}
\end{figure*}

The main results are shown in Fig.~\ref{fig:fits} and in Table~\ref{tab:fits}.
Figure~\ref{fig:fits} shows the observational RC (for simplicity we have
omited the error bars) as well as the fitted curves using equation~(\ref
{eq:modeltofit1}). Shown are also the individual contributions from luminous
matter, gas and the scalar field. It can be noted that the agreement is
quite good (within 5\% in all cases), which could have been expected since
there are three parameters to be adjusted. However, it should be noted that
this approach is made on a very solid theoretical basis, because we have
begun with a relativistic description of the galaxy. Moreover, the dark
matter in this model would be entirely constituted by the scalar field.\\

For the parameter $M/L$ our results are in very good agreement with previous
analyses which employ the dark-halo and the MOND approaches (e.g. \cite
{Begeman}). The remaining two parameters, $f_0$ and $b$, serve only to
determine completely the metric, and they do not have a direct physical
interpretation other than as part of the scalar field energy density. In
Table~\ref{tab:fits} the best fit parameters are listed, along with the
formal $1\sigma$ error in the fitted parameter.

\begin{table}
\caption{Sample of galaxies
\label{tab:sample}}
\begin{center}
\begin{tabular}{lcrrr}\hline

Galaxy & Type & Distance & Luminosity & $R_{HI}$ \\
       &      & (Mpc)    & ($10^9M_{\odot}$) & (kpc) \\\hline
NGC 2403         & Sc(s)III       &  3.25              &  7.90
            & 19.49 \\
NGC 2903         & Sc(s)I-II      &  6.40              & 15.30
            & 24.18 \\
NGC 6503         & Sc(s)II.8      &  5.94              &  4.80
            & 22.22 \\
NGC 3198         & Sc(rs)I-I      &  9.36              &  9.00
            & 29.92 \\
NGC 2841         & Sb             &  9.46              & 20.50
            & 42.63 \\
NGC 7331         & Sb(rs)I-I      & 14.90              & 54.00
            & 36.72 \\\hline
\end{tabular}
\end{center}
\end{table}

\begin{table}
\caption{Best-fit parameters \label{tab:fits}}
\begin{center}
\begin{tabular}{lrrrr}\hline

Galaxy & (M/L)$_{disk}$ & (M/L)$_{bulge}$ & b & f$_0$\\
        &               &                 & (kpc) & (kpc$^{-1}$)\\\hline
NGC 2403 	& 1.75   & --	& 1.63	& 0.0116 \\ 
	  	& 0.04	 & --	& 0.003	& $5.7\times 10^{-4}$ \\
NGC 2903 	& 2.98	 & --	& 8.33	& 0.0043 \\
	  	& 0.12	 & --	& 0.03	& $4.0\times 10^{-4}$ \\
NGC 6503 	& 2.12	 & --	& 1.79	& 0.013 \\
	 	& 0.09	 & -- 	& 0.01	& $8.7\times 10^{-4}$ \\
NGC 3198 	& 2.69	 & --	& 7.83	& 0.0054 \\
	  	& 0.08	 & --	& 0.02	& $3.0\times 10^{-4}$  \\
NGC 2841 	& 5.39   & 3.25	& 13.85 & 0.0039 \\
	 	& 0.34   & 0.36	& 0.16  & $2.5\times 10^{-4}$ \\
NGC 7331	& 5.06	 & 1.11 & 0.845	& 0.0013 \\
		& 0.23   & 0.06 & 0.002 & $8.9\times 10^{-5}$ \\\hline
\end{tabular}
\end{center}
\end{table}

\section{Conclusions.}

\label{sec:end}

In this work we have obtained an exact, axially symmetric and static
solution to the field equations of gravity coupled with a scalar 
field. This solution has been successfully employed as a model for a spiral
galaxy, and in particular, we have been able to reproduce the RC of matter
in these galaxies with an excellent agreement, both with the observations
themselves and with previous analyses of this kind of data.\\

It should be stressed that, although our model has three parameters to be
fitted, which in general allows for a great flexibility, they are obtained
from a theory that is of a fundamental nature, namely, the low energy limit
of a family of unification theories. This makes the calculations herein
shown to be natural since no {\it ad hoc} hypotheses are needed, in contrast
to the dark halo or the MOND models. We conclude that this work enables us
to state that scalar fields are strong candidates to constitute the dark
matter, not only at a cosmological scale, but also within spiral galaxies.

\acknowledgements{
We want to express our acknowledgment to the relativity
group in Jena for its kind hospitality and partial support.
This work was also partly supported by CONACyT, M\'exico.}

\appendix

\section{Some geometrical aspects of the metric.}

\label{app:geom}

In order to gain some insight into the solution (\ref{eq:metricBL}), we
shall consider a couple of issues related to the geometrical and
topological aspects of the metric. By defining: 
\[
x_{1}=\ln \sqrt{r^{2}+b^{2}\sin ^{2}\theta },
\]
\[
x_{2}=\arctan \left( \frac{\cot \theta }{\sqrt{1+b^{2}/r^{2}}}\right) ,
\]
\[
x_{3}=\varphi ,
\]
\[
x_{4}=f_{0}t,
\]
we have:

\begin{equation}  \label{eq:metricflat}
ds^2=\frac{1}{f_0r_0}e^{2x_1}(dx_1^2+dx_2^2+dx_3^2-c^2dx_4^2).
\end{equation}

\noindent From this, it can be argued that our coordinates ($\theta ,\varphi 
$) are not really angles, but in fact just cartesian coordinates. This,
however, can only be stated if we know the global topology of our
space-time, which we do not. As a counterexample, we can consider the
2-torus, ${\cal T}^{2}$; in this case, the metric is not only conformally
flat but in fact flat altogether, i.e., $ds^{2}=dx^{2}+dy^{2}$; with a
proper rescaling, this can be written $ds^{2}=d\theta _{1}^{2}+d\theta
_{2}^{2}$, where now $\theta _{1}$ and $\theta _{2}$ vary from $0$ to $2\pi $
and can be thought of effectively as angles. A similar example is the
2-sphere, ${\cal S}^{2}$. Therefore, our point of view is that, since
interpreting ($r,\theta ,\varphi $) as spherical-like coordinates allows us
to reproduce the rotation curves for galaxies, we can consider them in such
way, and in particular, $\varphi $ does represent angles about the axial
direction.

\section{Dark-halo profiles and the MOND hypothesis.}

\label{sec:dhmond}

The most commonly accepted approach to explain the RC is to assume that
there is some kind of unseen matter around the visible part of the galaxy,
forming what is usually called a `halo'; then, a mass density profile for
this dark matter is formulated and combined with data from the visible and
21 cm observations, and the model is fitted to the RC as obtained from
red-shifts in the galaxy. A quite broad family of density profiles is given
by (\cite{Zhao}):

\[
\rho_{halo}(r) = \frac{\rho_c}{(r/r_c)^\gamma[1+(r/r_c)^\alpha]^{(\beta-
\gamma)/\alpha}}, 
\]

\noindent where $\rho_0$ is the central density and $r_c$ is the `core'
radius, both of the halo. In particular, the simplest and most widely used
profile is the so called modified isothermal sphere (MIS), for which $%
(\alpha,\beta,\gamma)=(2,2,0)$. This model is attractive because there are
cosmological arguments which seem to suggest that, under the suitable
conditions, astronomical objects of this kind might actually evolve, the
dark matter being cold or hot depending on the evolutionary arguments (see,
e.g., \cite{Kravtsov} and references therein); however, a completely
satisfactory evolution scenario remains to be derived. In the actual fitting
procedure, it is usually more convenient to work with the asymptotic
circular velocity $v_h$ obtained from the isothermal sphere halo, by
applying the virial theorem and Newton's law:

\[
v_h = \sqrt{4\pi G\rho_0r_c^2}. 
\]

\noindent Beginning with the luminosity observations, $L(r)$, there are
three parameters to be adjusted (four in the case of galaxies with separate
observations from the disk and the bulge): the ratio $M/L$, usually assumed
to be constant over the whole optical disk, the core radius $r_c$ and the
asymptotic velocity $v_h$.\\

The other approach we shall consider here is known as the modified Newtonian
dynamics (MOND), which was proposed by Milgrom (1983); in this
case there is no dark matter at all, rather a deviation from the usual
Newton's second law would occur when one is dealing with very `small'
accelerations, where `small' means small with respect to some (in
principle, universal) critical acceleration parameter, $a_0$. Instead of
${\bf F}= m_g {\bf a}$, one would write:

\begin{equation}  \label{eq:mond}
{\bf g}_N = \mu(a/a_0){\bf a},
\end{equation}

\noindent where ${\bf g}_N$ is the conventional gravitational acceleration, $%
{\bf a}$ is the true acceleration of a particle with respect to some
fundamental frame ($a\equiv|{\bf {a}|}$), and $\mu$ is a function of $a/a_0$
of which only the asymptotic forms $\mu(a/a_0\gg1)\approx 1$,
$\mu(a/a_0\ll 1)\approx a/a_0$ are known. It can then be seen that for
accelerations much
larger than the acceleration parameter $a_0$, $\mu\approx 1$ and we
recover the
Newtonian dynamics. For the rotation law, the usual expression remains to be
valid: $v_C^2/r=a$, and also $g_N\approx MGr^{-2}$; combining this with
equation (\ref{eq:mond}), we get the asymptotic velocity:

\[
v_{C}^{4}=GMa_{0}. 
\]

\noindent In this case the acceleration parameter $a_0$ can be taken as
fixed so there is only one free parameter, again the ratio $M/L$.
Alternatively, $a_0$ can also be considered as a free parameter. Although
this approach works very well when fitting the rotation curves, there is no 
{\it a priori} reason to believe that a deviation of this kind could indeed
occur, so this is usually considered to be a purely phenomenological
description.

\end{document}